\input harvmac

\overfullrule=0pt
\abovedisplayskip=12pt plus 3pt minus 1pt
\belowdisplayskip=12pt plus 3pt minus 1pt
%
\def\tilde{\widetilde}

\def\to{\rightarrow}

\def\tC{{\tilde C}}

\def\bigone{\hbox{1\kern -.23em {\rm l}}}
\def\ZZ{\hbox{\zfont Z\kern-.4emZ}}
\def\half{{\litfont {1 \over 2}}}

\def\tr{{\rm tr}\,}

\def\Pf{{\rm Pf}\,}
\def\hA{{\hat A}}
\def\hS{{\hat S}}
\def\hF{{\hat F}}

\font\litfont=cmr6

\def\tD{{\tilde D}}

\def\tpa{{2\pi\alpha'}}
\def\boldS{{\bf S}}
\def\ola{\overleftarrow}
\def\ora{\overrightarrow}

\lref\noncommrefs{
A.~Connes, M.~R.~Douglas and A.~Schwarz,
{\it ``Noncommutative Geometry and Matrix Theory: Compactification 
on Tori''}, hep-th/9711162, JHEP {\bf 9802}, 003 (1998)\semi
M. Douglas and C. Hull, {\it ``D-branes and the 
Noncommutative Torus''}, hep-th/9711165, JHEP {\bf 9802}, 008 
(1998)\semi
F.~Ardalan, H.~Arfaei and M.~M.~Sheikh-Jabbari,
{\it ``Noncommutative Geometry from Strings and Branes''},
hep-th/9810072, JHEP {\bf 9902}, 016 (1999)\semi
C.~Chu and P.~Ho,
{\it ``Noncommutative Open String and D-brane''},
hep-th/9812219, Nucl.\ Phys.\ B {\bf 550}, 151 (1999)\semi
V. Schomerus, {\it ``D-branes and Deformation
Quantization''}, hep-th/9903205, JHEP {\bf 9906}, 030 (1999).}
\lref\myers{R. C. Myers, {\it ``Dielectric-Branes''}, hep-th/9910053,
JHEP {\bf 9912}, 022 (1999).}
\lref\ramswati{M. Van Raamsdonk and W. Taylor, {\it ``Multiple
Dp-branes in Weak Background Fields''}, hep-th/9910052,
Nucl. Phys. {\bf B573} (2000) 703.}
\lref\dharw{A.~Dhar and S.~R.~Wadia,
{\it ``A Note on Gauge Invariant Operators in 
Noncommutative Gauge Theories  and the Matrix Model''},
hep-th/0008144, Phys.\ Lett.\ B {\bf 495}, 413 (2000).}
\lref\watirev{W. Taylor, {\it ``The M(atrix) Model of M Theory''},
hep-th/0002016.}
\lref\seiwit{N. Seiberg and E. Witten, {\it ``String Theory and
Noncommutative Geometry''}, hep-th/9908142, JHEP {\bf 9909}, 032 (1999).}
\lref\seibnew{N. Seiberg, {\it ``A Note on Background
Independence in Noncommutative Gauge Theories, Matrix Model and
Tachyon Condensation''}, hep-th/0008013, JHEP {\bf 0009}, 003 (2000).}
\lref\twosens{B. Janssen and P. Meessen, {\it ``A Nonabelian
Chern-Simons Term for Non-BPS D-Branes''}, hep-th/0009025.}
\lref\garousitach{M.R. Garousi, {\it ``Tachyon Couplings on Non-BPS
D-Branes and Dirac-Born-Infeld Action''}, hep-th/0003122,
Nucl. Phys. {\bf B584} (2000) 284.}
\lref\smnvs{S. Mukhi and N.V. Suryanarayana, {\it ``Chern-Simons 
Terms on Noncommutative Branes''}, hep-th/0009101, 
JHEP {\bf 0011}, 006 (2000).}
\lref\tatar{R. Tatar, {\it ``T-Duality and Actions for Noncommutative
D-branes''}, hep-th/0011057.}
\lref\iikk{N. Ishibashi, S. Iso, H. Kawai and Y. Kitazawa, {\it
``Wilson Loops in Noncommutative Yang-Mills''},
hep-th/9910004, Nucl. Phys. {\bf B573}, 573 (2000).}
\lref\reyunge{S.-J. Rey and R. von Unge, {\it ``S-duality,
Noncritical Open String and Noncommutative Gauge Theory''},
hep-th/0007089, Phys. Lett. {\bf B499} 215 (2001).}
\lref\dasrey{S.R. Das and S.-J. Rey, {\it ``Open Wilson Lines
in Noncommutative Gauge Theory and Tomography of Holographic Dual
Supergravity''}, hep-th/0008042, Nucl. Phys. {\bf B590}, 453 (2000).}
\lref\ghi{D.J. Gross, A. Hashimoto and N. Itzhaki, {\it ``Observables 
of Noncommutative Gauge Theories''}, hep-th/0008075.}
\lref\garousi{M. Garousi, {\it ``Noncommutative World Volume
Interactions on D-branes and Dirac-Born-Infeld Action''},
hep-th/9909214, Nucl. Phys. {\bf B579} 209 (2000).}
\lref\mehenwise{T. Mehen and M. Wise, {\it ``Generalized $*$-products, 
Wilson Lines and the Solution of the Seiberg-Witten Equations''},
hep-th/0010204, JHEP {\bf 0012}, 008 (2000).}
\lref\micheliu{H. Liu and J. Michelson, {\it ``~$*$-Trek: The One
Loop N=4 Noncommutative SYM Action''}, hep-th/0008205.}
\lref\liu{H. Liu, {\it ``~$*$-Trek II: $*_n$ Operations, Open
Wilson Lines and the Seiberg-Witten Map''}, hep-th/0011125.}
\lref\dastrivedi{S.R. Das and S. Trivedi, {\it ``Supergravity
Couplings to Noncommutative Branes, Open Wilson Lines and Generalized
Star Products''}, hep-th/0011131, JHEP {\bf 0102}, 046 (2001).}
\lref\wyllard{N. Wyllard, {\it ``Derivative Corrections to D-brane 
Actions with Constant Background  Fields''}, hep-th/0008125,
Nucl. Phys. {\bf B598}, 247 (2001).}
\lref\kontsevich{M. Kontsevich, {\it ``Deformation Quantization of
Poisson Manifolds I''}, q-alg/9709040.}
\lref\catfeld{A.S. Cattaneo and G. Felder, {\it ``A Path Integral
Approach to the Kontsevich Quantization Formula''}, math.QA/9902090,
Comm. Math. Phys. {\bf 212} 591 (2000).}
\lref\jurcoschupp{B. Jurco and P. Schupp, {\it ``Noncommutative 
Yang-Mills from Equivalence of Star Products''},
hep-th/0001032, Eur. Phys. J. {\bf C14} 367 (2000).}
\lref\okawao{Y. Okawa and H. Ooguri, {\it ``An Exact Solution to 
Seiberg-Witten Equation of Noncommutative Gauge Theory''},
hep-th/0104036.}
\lref\smnvstwo{S. Mukhi and N.V. Suryanarayana, {\it ``Gauge-invariant 
Couplings of Noncommutative Branes to Ramond-Ramond Backgrounds''},
hep-th/0104045, JHEP {\bf 0105}, 023 (2001).}
\lref\micheliutwo{H. Liu and J. Michelson, {\it ``Ramond-Ramond 
Couplings of Noncommutative D-Branes''}, hep-th/0104139.}
\lref\garousi{M.~R.~Garousi,
{\it ``Non-commutative World-volume Interactions on D-branes and  
Dirac-Born-Infeld Action''}, hep-th/9909214,
Nucl.\ Phys.\ B {\bf 579}, 209 (2000).}
\lref\starprods{H.~Liu and J.~Michelson,
{\it ``*-TREK: The One Loop N = 4 Noncommutative SYM Action''},
hep-th/0008205\semi
F.~Ardalan and N.~Sadooghi,
{\it ``Anomaly and Nonplanar Diagrams in Noncommutative Gauge
Theories''}, hep-th/0009233\semi
A.~Santambrogio and D.~Zanon,
{\it ``One-loop Four-point Function in Noncommutative N=4 Yang-Mills 
Theory''}, hep-th/0010275, JHEP {\bf 0101}, 024 (2001)\semi
Y.~Kiem, D.~H.~Park and S.~Lee,
{\it ``Factorization and Generalized *-products''},
hep-th/0011233, Phys.\ Rev.\ D {\bf 63}, 126006 (2001)\semi
K.~Okuyama,
{\it ``Comments on open Wilson lines and generalized star products''},
hep-th/0101177, Phys.\ Lett.\ B {\bf 506}, 377 (2001).}
\lref\okawaoem{Y.~Okawa and H.~Ooguri,
{\it ``How Noncommutative Gauge Theories Couple to Gravity''},
hep-th/0012218,
Nucl.\ Phys.\ B {\bf 599}, 55 (2001).}
\lref\micheliuem{H.~Liu and J.~Michelson,
{\it ``Supergravity Couplings of Noncommutative D-branes''},
hep-th/0101016.}
\lref\earlyderivnc{
L.~Cornalba and R.~Schiappa,
{\it ``Matrix Theory Star Products from the Born-Infeld Action''},
hep-th/9907211\semi
Y.~Okawa,
{\it ``Derivative Corrections to Dirac-Born-Infeld Lagrangian and 
Non-commutative Gauge Theory''}, hep-th/9909132, 
Nucl.\ Phys.\ {\bf B566}, 348 (2000)\semi
L.~Cornalba,
{\it ``Corrections to the Abelian Born-Infeld Action Arising From  
Noncommutative Geometry''}, hep-th/9912293, JHEP {\bf 0009}, 017
(2000)\semi
S.~Terashima,
{\it ``On the Equivalence Between Noncommutative and Ordinary Gauge 
Theories''}, hep-th/0001111, JHEP {\bf 0002}, 029 (2000)\semi
Y.~Okawa and S.~Terashima,
{\it ``Constraints on Effective Lagrangian of D-branes from 
Non-commutative Gauge Theory''}, hep-th/0002194, 
Nucl.\ Phys.\ {\bf B584}, 329 (2000)\semi
L.~Cornalba,
{\it ``On the General Structure of the Non-Abelian Born-Infeld
Action''}, hep-th/0006018.}
\lref\earlyderivcomp{
J.H.~Schwarz, {\it ``Superstring Theory''}, Physics Reports {\bf 89},
223 (1982)\semi A.A. Tseytlin, {\it ``Vector Field Effective Actions
in the Open Superstring Theory''}, Nucl. Phys. {\bf B276}, 391
(1986)\semi O.D. Andreev and A.A. Tseytlin, {\it ``Partition Function
Representation for the Open Superstring Effective Action: Cancellation
of M\"obius Infinities and Derivative Corrections to the Born-Infeld
Lagrangian''}, Nucl. Phys. {\bf B311}, 205 (1988)\semi K. Hashimoto,
{\it ``Generalized Supersymmetric Boundary State''}, hep-th/9909095,
JHEP {\bf 0004}, 023 (2000).}

{\nopagenumbers
\Title{\vbox{
\hbox{hep-th/0106024}
\hbox{TIFR/TH/01-19}}}
{\centerline{Derivative Corrections from Noncommutativity}}
\centerline{Sumit R. Das,\footnote{}{E-mail: 
das@tifr.res.in, mukhi@tifr.res.in, nemani@tifr.res.in}
Sunil Mukhi and Nemani V. Suryanarayana}
\vskip 8pt
\centerline{\it Tata Institute of Fundamental Research,}
\centerline{\it Homi Bhabha Rd, Mumbai 400 005, India}

\vskip 2truecm
\centerline{\bf ABSTRACT}
\medskip
We show that an infinite subset of the higher-derivative $\alpha'$
corrections to the DBI and Chern-Simons actions of ordinary
commutative open-string theory can be determined using
noncommutativity. Our predictions are compared to some lowest order
$\alpha'$ corrections that have been computed explicitly by Wyllard
(hep-th/0008125), and shown to agree.
\vfill
\Date{May 2001}
\eject}
\ftno=0

\listtoc
\writetoc

\newsec{Introduction}

Noncommutativity has provided important new insights into the nature
of string theory\refs{\noncommrefs,\seiwit}. In the
presence of a 2-form $B$ field, one has the option to describe
effective actions for open strings in either commutative or
noncommutative descriptions. Using the continuous ``description
parameter'' $\Phi$ introduced in \refs\seiwit, one can actually
interpolate between the two types of descriptions, with $\Phi=B$
representing the commutative theory. The other useful choices are
$\Phi=0$ (which arises naturally in the point-splitting
regularization) and $\Phi=-B$ (which has been called the
``background-independent'' description\refs{\seiwit,\seibnew}, 
closely related to matrix theory). Depending on the question that one
wants to address, one can choose any of these descriptions for
convenience.

In different descriptions, the natural low-energy limits are also
different. The parameter governing higher-derivative corrections in
string theory is $\alpha'$, and one can take this to zero keeping
fixed various different quantities. An important limit in the
noncommutative description is the Seiberg-Witten limit,
$\alpha'\rightarrow 0$ keeping fixed the open-string metric $G$ along
with the 2-form field $B$ and the open-string coupling $G_s$. In this
limit, derivative corrections to the noncommutative actions (DBI and
Chern-Simons) vanish.

It was shown in Ref.\refs\seiwit\ that the noncommutative DBI action
is equivalent to the commutative one upto total derivative terms. In
recent times, it has been understood that exact equivalence (not just
upto total derivatives) of commutative and noncommutative actions can
be obtained if one additionally inserts an open Wilson
line\refs{\iikk,\dasrey,\ghi,\dharw} on the noncommutative side. This
has been useful in writing down gauge-invariant couplings of open
string modes to closed-string NSNS fields\refs{\liu,\dastrivedi} and
RR fields\refs{\smnvs,\okawao,\smnvstwo,\micheliutwo}. This has
provided a powerful tool to extract new information. For example it
allows one to obtain an exact expression for the Seiberg-Witten map
between commutative and noncommutative gauge couplings in the abelian
case.

In this note we exhibit a new application of the noncommutative
description of string theory. We start by assuming exact equivalence
of the commutative and noncommutative actions, including all
derivative corrections on both sides. It is important to note that
terms which, for constant backgrounds, would have been total
derivatives, are also retained. Next we take the Seiberg-Witten limit,
which sets to zero the $\alpha'$ corrections on the noncommmutative
side, and reduces it to a sum of Yang-Mills and Chern-Simons
actions. Comparison of the two sides now yields definite predictions
for the derivative corrections on the commutative side, or at least
those corrections (there are infinitely many) that survive the
Seiberg-Witten limit.

Earlier attempts to study derivative corrections to the DBI action
using noncommutativity can be found in Ref.\refs\earlyderivnc. The
principal new ingredient in our work is the fact that with open Wilson
lines, one has exact agreement between commutative and noncommutative
actions, including couplings to closed-string backgrounds at nonzero
momentum. This allows us to make very explicit predictions and compare
them with perturbative string amplitudes.

Some of the derivative corrections in open-string theory were computed
explicitly in recent times\refs\wyllard\ in the boundary-state
formalism\foot{Earlier work on the computation of such corrections can
be found in Ref.\refs\earlyderivcomp. We use the results of
Ref.\refs\wyllard\ as they are the most complete to date, and also
because the choice of field variables turns out to have a special
significance, as we will see.}. We will show in a number of cases that
the numerical coefficients and index structures given by these
computations can be reproduced using our arguments, by expanding the
$n$-ary product $*_n$ that has recently played an important role in
the noncommutative description of string effective actions.

The agreement between the predictions of noncommutativity and the
computations of Ref.\refs\wyllard\ might seem somewhat fortuitous,
given that there is always a freedom of field redefinitions. We will
comment on this point in some detail in the Conclusions.

\newsec{Background and Proposal}

In what follows, we will always work with the BPS D9-branes of type IIB
string theory, though the discussion can in principle be extended to
lower D-branes. The DBI and Chern-Simons actions on the brane in the
commutative description are:
\eqn\commaction{
\eqalign{
S_{DBI} &= {1\over g_s} \int \sqrt{g+ \tpa(B+F)}\cr 
S_{CS} &= {1\over g_s} \int \sum_n C^{(n)}\, e^{\tpa(B+F)}}}
In the latter expression, the exponential is to be expanded to keep
the 10-form part.

The noncommutative description\refs\seiwit\ is parametrized by 
the noncommutativity parameter $\theta$, the open-string metric 
$G_{ij}$, the open-string coupling $G_s$, and a ``description
parameter'' $\Phi$, in terms of which the relationship between 
closed-string and open-string parameters is given by:
\eqn\ocreln{
\eqalign{
N^{ij}\equiv \left({1\over g+\tpa B}\right)^{ij} &= {\theta\over\tpa}
+ {1\over G+\tpa \Phi}\cr &\cr
{\sqrt{\det(g+\tpa B)}\over g_s} &= {\sqrt{\det(G+\tpa\Phi)}\over
G_s}\cr} }
In what follows, it is most convenient to work in the $\Phi=-B$
description, where the contact with matrix theory is explicit. In this
description, the DBI action can be written equivalently in two
convenient forms:
\eqn\ncdbi{
\eqalign{
\hS_{DBI} &= {1\over G_s} \int \sqrt{G + 2\pi\alpha'(\hF-B)}\cr
&= {1\over g_s}\int {\Pf Q\over \Pf\theta}
\sqrt{g + 2\pi\alpha'\,Q^{-1}}\cr} }
Here, $G_s$, $G$ and $\theta$ are the open-string coupling, metric and
noncommutativity parameter respectively, defined by
\eqn\openstr{
\theta^{ij} = (B^{-1})^{ij},\qquad
G_{ij} = - (\tpa)^2 B_{ik}\,g^{kl}\,B_{lj},\qquad
G_s = g_s \sqrt{\det \tpa B\over \det g} }
while $Q^{ij}$ is given by
\eqn\qij{
Q^{ij} = \theta^{ij} - \theta^{ik}\hF_{kl}\,\theta^{lj} }
and $\Pf$ denotes the Pfaffian or square root of the determinant. We
also  note the following explicit expression for $Q^{-1}$, which will
be useful later on:
\eqn\qinv{
Q^{-1} = \theta^{-1} + \hF{1\over 1-\theta\hF} }

As is well-known, the above relations also hold in the Seiberg-Witten
limit\refs\seiwit:
\eqn\seiwitlim{
\alpha'\to 0,\qquad G,B, G_s~~{\rm fixed}}
regardless of the description parameter $\Phi$. Since we will be
mainly working in this limit in what follows, our results can also be
interpreted as being valid in any description. In particular, this
observation explains the agreement of the $\Phi=-B$ results of
Refs.\refs\dastrivedi\ with the explicit string amplitude calculations
of Refs.\refs{\okawaoem,\micheliuem}. The latter of course involve the
point-splitting regularization and therefore they correspond to the
$\Phi=0$ description.

The noncommutative Chern-Simons action for constant fields
can be written as follows\refs\smnvs\
\eqn\nccs{
\hS_{CS} = {1\over g_s}\int {\Pf Q\over \Pf\theta}\sum_n C^{(n)}
e^{\tpa Q^{-1}} }
where the exponential is to be expanded so that the total form has the
rank of the brane worldvolume, namely 10 in our case. 

For nonconstant fields, the above actions are not gauge-invariant and
one needs to introduce an open Wilson line. This is defined as
\eqn\wilsonline{
W(x,C) \equiv \exp\left(- i \int_0^1 d\tau {\del\xi^i(\tau)\over
\del\tau}\hA_i\Big(x+ \xi(\tau)\Big) \right) }
where the contour of the Wilson line is defined in terms of a fixed
momentum $k$ by $\xi^i(\tau) = \theta^{ij}k_j\,\tau$ with $0\le\tau\le
1$. This operator must be inserted to make the coupling to a
closed-string mode of momentum $k$ gauge-invariant.  For example, if
we consider the linearized coupling to a dilaton $\tD(k)$, the DBI
action must be replaced by\refs{\liu,\dastrivedi}
\eqn\wilsondbi{
\hS_{DBI}(k) = {\tD(-k)\over g_s} \int L_*\left\{ {\Pf Q\over \Pf\theta}
\sqrt{g + 2\pi\alpha'\,Q^{-1}}\,W(k,C)\right\}* e^{ik.x} }
The operation $L_*$ consists of smearing all operators along the
contour of the Wilson line and path-ordering the resulting expression
with respect to the noncommutative $*$ product.

In the same way, the coupling of open-string modes to a Ramond-Ramond
form of nonzero momentum is given by\refs{\okawao,\smnvstwo,\micheliutwo}
\eqn\wilsoncs{
\sum_n {\tilde C}^{(n)}(-k)\int
~L_* \left\{{\Pf Q\over \Pf\theta}~e^{\tpa Q^{-1}}\, W(x,C)\right\} 
* e^{ik.x}}
where again we need to pick out the 10-form contributions in the above
expression.

As explained in Refs.\refs{\mehenwise,\liu}, the expansion 
of the above expressions can be written in terms of an $n$-ary
product called $*_n$, which maps a collection of $n$ functions
$f_1,f_2,\ldots,f_n$ to a single function that we denote $\langle f_1,
f_2,\ldots,f_n\rangle_{*_n}$. The definition of $*_2$ is relatively
simple: 
\eqn\startwodef{
\langle f(x),g(x) \rangle_{*_2} \equiv 
f(x){\sin(\half\ola{~\partial_p}\,\theta^{pq}\ora{\,\partial_q\,})
\over \half\ola{~\partial_p}\,\theta^{pq}\ora{\,\partial_q\,} } g(x) }
More information about the role of $*_n$ products, and 
general formulae, can be found in
Refs.\refs{\garousi,\mehenwise,\liu,\dastrivedi,\starprods}. 

More specifically, expanding out an $L_*$ product leads to the Fourier
transform of the $*_n$ products. At this point it is often more
convenient to go back to position space. Hence in what follows, we
will usually work in position space, but will be forced to use
momentum space whenever the $L_*$ product is yet to be expanded
out. We hope this will be clear from the context.

Now let us summarize the basic approach of this paper.  The DBI and CS
actions written here will in general have corrections that involve
higher powers of $\alpha'$. Let us denote these corrections by
$\Delta\hS_{DBI}$ and $\Delta\hS_{CS}$ respectively.  The requirement
that noncommutative and commutative actions are really the same means
that
\eqn\equivaction{
\eqalign{
S_{DBI} + \Delta S_{DBI} &= \hS_{DBI} + \Delta \hS_{DBI}\cr
S_{CS} + \Delta S_{CS} &= \hS_{CS} + \Delta \hS_{CS}\cr }}
Here the terms on the left hand side are the open-string effective
actions plus their derivative corrections in the usual commutative
description.

Note that in $\Delta\hS_{DBI}$ and $\Delta\hS_{CS}$, indices are
always contracted with the open string metric $G_{ij}$. Therefore in
the Seiberg-Witten limit all these noncommutative corrections vanish,
and the identities in Eq.\equivaction\ reduce to
\eqn\swlimaction{
\eqalign{
S_{DBI}\Big|_{SW} + \Delta S_{DBI}\Big|_{SW} &= \hS_{DBI}\Big|_{SW} \cr
S_{CS}\Big|_{SW} + \Delta S_{CS}\Big|_{SW} &= \hS_{CS}\Big|_{SW} }}
where the subscript $SW$ indicates that the Seiberg-Witten limit 
has been taken.

In what follows, our strategy will be to derive information about
$\Delta S_{DBI}\Big|_{SW}$ and $\Delta S_{CS}\Big|_{SW}$ using the
exact knowledge of the commutative and noncommutative DBI and
Chern-Simons actions. Some terms in $\Delta S_{DBI}$ and $\Delta
S_{CS}$ have been computed in Ref.\refs\wyllard\ and we will compare
them in the SW limit with the prediction from the RHS, finding
complete agreement. We will discuss to what extent this allows us to
recover full information about these terms away from the SW limit.

\newsec{The Dirac-Born-Infeld Action}

In this section we wish to compare the sum of the commutative DBI
action $S_{DBI}$ plus the derivative corrections to it $\Delta
S_{DBI}$ (some of which are computed in Ref.\refs\wyllard) with the
noncommutative DBI action $\hS_{DBI}$, after taking the Seiberg-Witten
limit on both sides. 

\subsec{Dilaton Coupling, Order $F^2$}

The dilaton couples to the entire Lagrangian density, so we need to
consider the full DBI action. We will start by restricting to terms
quadratic in $F$. To this order, we have:
\eqn\dbismallf{
\eqalign{
S_{DBI} = \int {\sqrt{\det (g+\tpa B)}\over g_s}\Big[&
1 + {\tpa\over 2} \tr (NF) - {(\tpa)^2\over 4} \tr(NFNF)\cr
&+ {(\tpa)^2\over 8}\big(\tr NF\big)^2 + \ldots \Big]\cr} }
In the Seiberg-Witten limit we have $N^{ij}\to {\theta^{ij}\over \tpa}$ 
and therefore:
\eqn\sdbisw{
S_{DBI}\Big|_{SW} = \int {\sqrt{\det (g+\tpa B)}\over g_s}\Big[
1 + {1\over 2} \tr (\theta F) - {1\over 4} \tr(\theta F\theta F)
+ {1\over 8}\big(\tr \theta F\big)^2 + \ldots \Big] }
Note that, here and later in the paper, we insert this limit only in
the bracketed series expansion, leaving the prefactor untouched. This
is because the prefactor will eventually cancel with the corresponding
prefactor on the noncommutative side when we compare the two.

Let us now convert the commutative field strengths $F$ appearing in
this expression into noncommutative field strengths $\hF$, using the
Seiberg-Witten map. To the order that we need it, this map is:
\eqn\swmaplow{
F_{ab} = \hF_{ab} + \theta^{kl}\left(
\langle \hA_k,\del_l \hF_{ab}\rangle_{*_2} 
- \langle \hF_{ak},\hF_{bl} \rangle_{*_2}\right) }
where
\eqn\fhatdef{
\hF_{ab} = \del_a \hA_b - \del_b \hA_a + \theta^{kl}\langle
\del_k \hA_a,\del_l \hA_b\rangle_{*_2} }
Here we have used an identity relating the Moyal $*$ commutator and
the $*_2$ product:
\eqn\starid{
-i[f,g]_* = \theta^{ij} \langle \del_i f,\del_j g\rangle_{*_2} }
Inserting the Seiberg-Witten map into Eq.\sdbisw, we find
\eqn\sdbiswsw{
\eqalign{
S_{DBI}\Big|_{SW} =~& \int {\sqrt{\det (g+\tpa B)}\over g_s}\Big[
1 + \theta^{ij}\del_j \hA_i + \half \theta^{ba}\theta^{kl}\langle
\del_k \hA_a,\del_l \hA_b\rangle_{*_2} \cr 
&+\half \theta^{ab} \theta^{kl} \Big( \langle \hA_k, \del_l
\hF_{ab}\rangle_{*_2} - \langle \hF_{ak},\hF_{bl}\rangle_{*_2}\Big)
- {1\over 4} \theta^{ij}\theta^{kl} \hF_{jk} \hF_{li}\cr
&+ {1\over 8} \Big(\theta^{ij}\hF_{ij}\Big)^2 \Big] \cr}}
Some manipulation of the last few terms permits us to rewrite this as: 
\eqn\scommdbi{
\eqalign{
S_{DBI}\Big|_{SW} =~& \int {\sqrt{\det (g+\tpa B)}\over g_s}\Big[
1 + \theta^{ij}\del_j \hA_i + \half \theta^{ij}\theta^{kl}\langle
\del_j \hA_k,\del_l \hA_i\rangle_{*_2} \cr 
&+\half \theta^{ba} \theta^{kl} \langle \hA_k, \del_l
\hF_{ab}\rangle_{*_2} 
+ {1\over 4} \theta^{ij}\theta^{kl}\Big(
\langle\hF_{jk}, \hF_{li}\rangle_{*_2} - \hF_{jk} \hF_{li} \Big)\cr
&+ {1\over 8} \theta^{ij}\theta^{kl}\langle
\hF_{ji},\hF_{lk}\rangle_{*_2} 
- {1\over 8}  \theta^{ij}\theta^{kl}\Big(
\langle\hF_{ji}, \hF_{lk}\rangle_{*_2} - \hF_{ji} \hF_{lk} \Big) \Big]}}
which is the form in which it will be useful.

Let us now turn to the noncommutative side. Here, we only need to keep
the terms arising from expansion of the Wilson line, since all other
terms are suppressed by powers of $\alpha'$ in the Seiberg-Witten
limit. The Wilson line gives us:
\eqn\wilsonexp{
\hS_{DBI}\Big|_{SW} = \int {\sqrt{\det (G+\tpa \Phi)}\over G_s}
\Big[1 + \theta^{ij}\del_j \hA_i + \half
\theta^{ij}\theta^{kl}
\del_j \del_l \langle \hA_i,\hA_k\rangle_{*_2}\Big] }
After some rearrangements of terms, this can be written:
\eqn\sncdbi{
\eqalign{
\hS_{DBI}\Big|_{SW} =~& \int {\sqrt{\det (G+\tpa \Phi)}\over G_s}\Bigg[
1 + \theta^{ij}\del_j \hA_i + \half \theta^{ij}\theta^{kl}\langle
\del_j \hA_k,\del_l \hA_i\rangle_{*_2} \cr
&+\half \theta^{ba} \theta^{kl} \langle \hA_k, \del_l
\hF_{ab}\rangle_{*_2} 
+ {1\over 8} \theta^{ij}\theta^{kl}\langle
\hF_{ji},\hF_{lk}\rangle_{*_2} 
\Bigg]}}
Now we can take the difference of Eqs.\sncdbi\ and \scommdbi. The
prefactor in front of each expression is the same, by virtue of 
Eq.\ocreln. Apart from this factor and the integral sign, the result is:
\eqn\finaldiff{
\hS_{DBI}\Big|_{SW} - S_{DBI}\Big|_{SW}= 
{1\over 4} \theta^{ij}\theta^{kl}
\Big(
\langle\hF_{jk}, \hF_{li}\rangle_{*_2} - \hF_{jk} \hF_{li} \Big)
- {1\over 8}  \theta^{ij}\theta^{kl}\Big(
\langle\hF_{ji}, \hF_{lk}\rangle_{*_2} - \hF_{ji} \hF_{lk} \Big) }
To the order in which we are working, we may replace $\hF$ by $F$ 
everywhere in this expression.

This, then, is our prediction for the correction $\Delta
S_{DBI}$, to order $(\alpha')^2$ and to quadratic order in
the field strength $F$, after taking the Seiberg-Witten limit. We note
that this is manifestly a higher-derivative correction: it vanishes
for constant $F$, for which the $*_2$ product reduces to the ordinary
product. Expanding the $*_2$ product to 4-derivative order, we find
that
\eqn\dbipred{
\eqalign{
\Delta S_{DBI}\Big|_{SW} = -{1\over 96} \Big[&
\theta^{ij} \theta^{kl} \theta^{mn} \theta^{rs} \,
\del_m \del_r F_{jk}\, \del_n \del_s F_{li}\cr 
& -\half \theta^{ij} \theta^{kl} \theta^{mn} \theta^{rs} \,
\del_m \del_r F_{ji}\, \del_n \del_s F_{lk} \Big] \cr}} 
This prediction may now be compared with the computation reported in
Eq.(4.1) of Ref.\refs\wyllard, which gives:
\eqn\wyllarddbi{
\eqalign{
\Delta S_{DBI}\Big|_{SW} = -{(\tpa)^4\over 96} \Big[&
h^{ij} h^{kl} h^{mn} h^{rs} \,
\del_m \del_r F_{jk}\, \del_n \del_s F_{li}\cr 
& -\half h^{ij} h^{kl} h^{mn} h^{rs} \,
\del_m \del_r F_{ji}\, \del_n \del_s F_{lk} \Big] \cr}} 
where the matrix $h^{ij}$ is defined as:
\eqn\hdef{
h^{ij}\equiv \left({1\over g+\tpa (B+F)}\right)^{ij} }
Taking the Seiberg-Witten limit, which amounts to the replacement
$\tpa h\to (1+\theta F)^{-1}\theta$, and further
restricting to terms quadratic in $F$, we find exact agreement with
Eq.\dbipred\ above.

\subsec{Graviton Coupling, Order $F^2$}

In this subsection, we will compare the coupling of the bulk graviton
to the energy-momentum tensor on the commutative and noncommutative
sides. On the commutative side, we start again with the expression 
in Eq.\dbismallf, but this time we use the full form of $N$ as defined 
in Eq.\ocreln:
\eqn\nrep{
N^{ij}\equiv \left({1\over g+\tpa (B+F)}\right)^{ij} =
{\theta^{ij}\over \tpa} + M^{ij} }
where
\eqn\mdef{
M^{ij} \equiv \left({1\over G+\tpa \Phi}\right)^{ij} }
As the linear coupling to the graviton starts at order $(\alpha')^2$,
we now have to go beyond the leading term in the Seiberg-Witten
limit. Hence we will keep terms up to order $M^2$.

Expanding $S_{DBI}$ around this limit and keeping terms to order
$(\alpha')^2$, and using the Seiberg-Witten map, we find:
\eqn\sdbigrav{
\eqalign{
&S_{DBI} = \int {\sqrt{\det(g+\tpa B)}\over g_s}\Bigg[1 + {\tpa\over 2}
\Big\{M^{ji} (F_{ij} + \theta^{kl}\langle A_k,\del_l
F_{ij}\rangle_{*_2} + \theta^{kl}\langle F_{jk}, F_{li}\rangle_{*_2} )
\Big\} \cr
&+ {(\tpa)^2\over 8} \Big\{ {2\over\tpa}(\tr MF)(\tr \theta F)
+ (\tr MF)^2\Big\}  - {(\tpa)^2\over 4} \Big\{ \tr MFMF + {2\over
\tpa} \tr MF\theta F\Big\}\cr
& + {\rm terms~not~involving~}M + {\rm order}~F^3 \Bigg] \cr}}

Turning now to the noncommutative action, the graviton coupling is
obtained by expanding the DBI action around the Seiberg-Witten limit
to order $(\alpha')^2$. There could in principle have been other
relevant $\alpha'$ corrections to the DBI action, but these are absent 
by virtue of the result in Refs.\refs{\okawaoem,\micheliuem} that the
energy-momentum tensor as calculated from string amplitudes agrees
with the one obtained by just expanding the DBI action to this
order. Thus we have, in momentum space:
\eqn\hsdbigrav{
\eqalign{
\hS_{DBI} &= {1\over G_s}\int L_*\left[\sqrt{\det(G+\tpa(\hF+\Phi))}\,
W(x,C)\right] * e^{ik.x} \cr &= {1\over G_s}\int \sqrt{\det G}\, L_*
\left[\Bigg( 1- {1\over 4}(\tpa)^2 \tr G^{-1}(\hF +
\Phi) G^{-1} (\hF + \Phi)\Bigg)\,W(x,C)\right]
\star e^{ik.x}\cr
&+\ldots}}
The piece of the above expression that is order 1 in $\alpha'$ has
already been computed earlier for the dilaton coupling. It contributes 
to the coupling of the trace of the graviton. The new nontrivial
coupling is given by the order $(\alpha')^2$ term.

To compare with the commutative side, it is convenient to expand the
above action differently, in terms of $M$ rather than $G$. We get:
\eqn\hsdbidiff{
\eqalign{
\hS_{DBI} &= {1\over G_s}\int {\sqrt{\det (G+\tpa\Phi)}}
\Bigg[L_*(W(x,C)) +
{\tpa\over 2}\Big\{\tr MF + M^{kl}\theta^{ij}\langle \del_j F_{lk},
A_i\rangle_{*_2} \cr 
&+ \half\langle\tr MF,\tr\theta
F\rangle_{*_2}\Big\}
- {(\tpa)^2\over 4}\tr \langle MF,MF\rangle_{*_2}
+ {(\tpa)^2\over 8}\langle \tr MF,\tr MF\rangle_{*_2} + \ldots
\Bigg]}}
Now taking the difference of the noncommutative and commutative
actions in Eqs.\hsdbidiff\ and \sdbigrav, and expanding the result to
4-derivative order, we get the prediction:
\eqn\gravdelta{
\eqalign{
\Delta & S_{DBI}\Big|_{SW} = - {\tpa\over 48}\Big\{
M^{ij}\theta^{kl}\theta^{mn}\theta^{rs}\del_m \del_r F_{jk}
\del_n\del_s F_{li} - \half
M^{ij}\theta^{kl}\theta^{mn}\theta^{rs}\del_m\del_r F_{ji}
\del_n\del_s F_{lk}\Big\}\cr
&- {(\tpa)^2\over 96}\Big\{
M^{ij}M^{kl}\theta^{mn}\theta^{rs}\del_m \del_r F_{jk}
\del_n\del_s F_{li} - \half
M^{ij}M^{kl}\theta^{mn}\theta^{rs}\del_m\del_r F_{ji}
\del_n\del_s F_{lk}\Big\}\cr}}
Note that contrary to appearances, both of the above terms are of
order $(\alpha')^2$. This is because if one inserts $G$ in place of $M$ in 
the first line, the result vanishes.

The above can now be compared with the result of Ref.\refs\wyllard\
quoted above in Eq.\wyllarddbi. Here one has to insert
$h\sim{\theta\over\tpa} + M$ which is true after we neglect the $F$ in
the denominator of $h$ (see Eqs.\hdef,\ocreln).  We see that
Eq.\gravdelta\ is reproduced perfectly if one retains only the term
proportional to $\theta^{mn}\theta^{rs}$ from $h^{mn}h^{rs}$, while
keeping the terms proportional to $M^{ij}\theta^{kl}$ and $M^{ij}
M^{kl}$ in $h^{ij}h^{kl}$.

One can, however, keep factors of $M$ in the expansion of
$h^{mn}h^{rs}$, and this leads to other terms from Eq.\wyllarddbi\
that are not reproduced by our computations. These terms are
comparable to curvature couplings in that they are linear or quadratic
in the metric, and quadratic in derivatives. Since the computations of
Ref.\wyllard\ were performed in flat space neglecting the presence of
curvature couplings, it is perhaps not surprising that we do not find
agreement for those terms. We hope to return to this point in the
future.

\newsec{Chern-Simons Action}

In this section we compare the Chern-Simons actions in the commutative
and noncommutative descriptions. The first such comparison is that of
the coupling to the 10-form RR potential $C^{(10)}$. In this case we
have, in momentum space,
\eqn\tenform{
\eqalign{
S_{CS} &= {1\over g_s}\, \tC^{(10)}(-k)\, \delta(k)\cr
\hS_{CS} &= {1\over g_s}\, \tC^{(10)}(-k) \int 
L_* \left\{{\Pf Q\over \Pf\theta}\, W(x,C)\right\} 
* e^{ik.x}}}
In this case, it has been argued\refs\wyllard\ that $\Delta S_{CS}=0$,
so the two expressions above agree exactly, leading to the topological
identity of Refs.\refs{\okawao,\smnvstwo,\micheliutwo}. In these
papers it was also shown that an analogous result holds for comparison
of the coupling to the 8-form RR potential $C^{(8)}$, leading to an
exact expression for the Seiberg-Witten map in the abelian case.

For the coupling to the RR forms $C^{(6)},C^{(4)},C^{(2)}$ and
$C^{(0)}$, there are in general $\alpha'$ corrections involving
derivatives of the field strength. A subset of these has been computed 
explicitly in Ref.\refs\wyllard. We will parametrize these derivative
corrections as follows:
\eqn\csparam{
S_{CS} + \Delta S_{CS} = {1\over g_s}\int \sum_n C^{(n)}\wedge
e^{\tpa(B+F)} \wedge e^{W_4 + W_6 + W_8 + W_{10}} }
where $W_{2n}$ are $2n$-forms made out of $F$ and its derivatives,
containing explicit powers of $\alpha'$. The expression on the RHS is
to be expanded and then the forms of total dimension 10 are kept.
This parametrization is inspired by the lowest order computations in
Ref.\refs\wyllard, which we will confirm using noncommutativity, and
which give rise to rather simple expressions for the
$W_{2n}$. However, it is important to keep in mind that Eq.\csparam\
is a general parametrization. The results that one finds for
derivative corrections can always be cast in this form. A point to
note here is that our notation is not identical to that of
Ref.\refs\wyllard, for example what is called $W_8$ there is the sum
of our $W_8$ and $\half W_4\wedge W_4$.

\subsec{4-Form Corrections, Order $F^2$}

We turn now to the 4-form that couples to $C^{(6)}$. The commutative
Chern-Simons coupling in this case is proportional to
$(B+F)\wedge(B+F)$. Expanding this leads to three terms, proportional
to $B\wedge B$, $B\wedge F$ and $F\wedge F$. It is easy to see that on
the noncommutative side too there are three terms\refs\smnvs, which
can be respectively matched with these three. Matching of $B\wedge B$
leads to the topological identity which was already discovered by
examining the RR 10-form coupling. Matching $B\wedge F$ similarly
leads to the Seiberg-Witten map. Hence the only new information comes
from matching $F\wedge F$, from which we will learn about derivative
corrections. This pattern will be repeated when we study lower RR
forms. Therefore at each stage, it suffices to examine the $F^n$ part of
the CS coupling.

Hence the Chern-Simons coupling that we will now study (an overall
factor of $(\tpa)^2$ has been removed) is
\eqn\csfourbasic{
S_{CS}= {1\over g_s}\,\int C^{(6)}\wedge \left(\half F\wedge F\right)}

According to our parametrization, the correction $\Delta S_{CS}$ 
is of the form:
\eqn\wfourcorr{
\Delta S_{CS} = {1\over g_s}\,\int C^{(6)}\wedge W_4}
where $W_4$ is a 4-form. This was computed to 4-derivative order,
or equivalently order $(\alpha')^2$, in Ref.\refs\wyllard, where it
was found to be:
\eqn\wfour{
W_4 = (\tpa)^2\,{\zeta(2)
\over 8\pi^2}\, \tr (h\boldS \wedge h\boldS) +\ldots}
The 2-form $\boldS_{ij}$ in the above expression is defined by
\eqn\twoforms{
\eqalign{
\boldS_{ij} &\equiv \half S_{ij,ab}\, dx^a\wedge dx^b\cr
&\equiv \half\left(\del_i \del_j F_{ab} + (\tpa)\,2h^{cd}\,\del_i
F_{ac}\,\del_j F_{db} \right) dx^a\wedge dx^b}}
and contractions are carried out using $h$, defined in Eq.\hdef\ above.

In the Seiberg-Witten limit, $h^{ij}\to {\theta\over\tpa}$ and this
correction becomes
\eqn\fourderivsw{
\Delta S_{CS}\Big|_{SW} = {1\over g_s}\,\int C^{(6)}
\wedge {1\over 48}
\tr (\theta\boldS \wedge \theta\boldS) }
where we have inserted $\zeta(2)={\pi^2\over 6}$.

Now let us work in the limit of small field strength, keeping only the
leading (in this case quadratic) terms as $F\to 0$, and test whether
Eq.\wilsoncs\ indeed reproduces this term. For the 4-form correction,
the operative term in Eq.\wilsoncs\ (again with an overall $(\tpa)^2$
removed) is:
\eqn\wilsoncsoper{
\half\int L_*\left[ \Pf(1-\theta\hF)\, 
\Bigg(\hF{1\over 1-\theta\hF}\Bigg)\wedge
\Bigg(\hF{1\over 1-\theta\hF}\Bigg)\,
W(x,C) \right] * e^{ik.x} }
Since we are working to order $F^2$, we can neglect the difference
between $F$ and $\hF$, and also the effect of the Pfaffian, the
$(1-\theta\hF)$ denominators, and the Wilson line. Indeed, the only
effect of noncommutativity that we need to keep is the fact that the
$L_*$ prescription leads to $*_n$ products, in this case $*_2$. Thus
the above expression reduces to 
\eqn\wilsonred{\half \int \langle F\wedge
F\rangle_{*_2}\, e^{ik.x}}
In the small-$F$ limit we can also reduce the 2-form $\boldS$ in
Eq.\twoforms\ to
\eqn\sreduces{
S_{ij,ab} \sim
\del_i \del_j F_{ab} }
As a result, Eq.\swlimaction\ tells us that we should find:
\eqn\smallf{
\half F\wedge F + {1\over 48}\tr (\theta\boldS \wedge \theta\boldS)
=\half\langle F \wedge F\rangle_{*_2}}
(for the $*_n$ product of $n$ differential forms, we use the notation
$\langle f_1\wedge f_2\wedge\ldots f_n\rangle_{*_n}$). 

It is easy to check, from the definition of the $*_2$ product in
Eqn.\startwodef, that:
\eqn\easycheck{
\half\langle F\wedge F\rangle_{*_2}= \half F\wedge F + {1\over 48}
\tr (\theta^{ij}\del_j\del_k F \wedge \theta^{kl}\del_l\del_m F)
 + \ldots}
in agreement with the LHS of Eq.\smallf.

In this discussion of 4-form corrections, we have so far restricted
our attention to 4-derivative terms that are quadratic in $F$. Let
us now go beyond the 4-derivative approximation but retain the
restriction to quadratic $F$ (thus, $F$ is small but not slowly
varying). In this case, following the techniques of Ref.\refs\wyllard,
one could explicitly compute higher-order corrections in $\alpha'$ to
Eq.\wfour. This has not been actually done so far, to the best of our
knowledge. But from our considerations, we can predict what the result
will be in the Seiberg-Witten limit, to every order in derivatives!
Indeed, our prediction amounts to the statement that to quadratic
order in $F$,
\eqn\allderiv{
\Delta S_{CS}\Big|_{SW} = {1\over g_s}\int C^{(6)}\wedge\left\{
\half \langle F\wedge F\rangle_{*_2} - \half F\wedge F \right\} }
where the RHS has infinitely many higher-derivative terms.

For example, the 8-derivative correction arising out of this is:
\eqn\eightderiv{
\Delta S_{CS}\Big|_{SW} = {1\over g_s}\int C^{(6)}\wedge\left\{
{1\over 3840}\, \theta^{ij} \theta^{kl} \theta^{mn} \theta^{pq}\,
\del_i  \del_k  \del_m  \del_p F \wedge
\del_j  \del_l  \del_n  \del_q F \right\} }
and this should be checked by explicit computation of string
amplitudes.

It is tempting to speculate that one can read off the result even away
from the Seiberg-Witten limit, by making the substitution
\eqn\thetasub{
\theta^{ij} \to \tpa N^{ij}= \tpa\left( {1\over g+ \tpa B} \right)^{ij} }
The problem is that this substitution is not unique. The LHS is
antisymmetric, so there could be terms that are nonvanishing in
general but vanish in the SW limit. If so, we would not find them by
our procedure. Nevertheless, if the above substitution turns out to make
sense, it would amount to saying that the 4-form corrections to all
derivative orders, but quadratic in $F$, are encoded in a $*_2$ product
whose noncommutativity parameter is $\tpa h$ (a matrix of no definite
symmetry) rather than $\theta$. This is suggestive of a beautiful
mathematical structure underlying stringy $\alpha'$ corrections.

\subsec{6-Form Corrections, Order $F^3$}

Let us now look at corrections to the 6-form that couples to the RR
4-form potential $C^{(4)}$. We continue to work in the limit of small
$F$, so we only keep the lowest power of $F$, in this case $F^3$, in
all terms.  The basic Chern-Simons coupling of interest in this
subsection is:
\eqn\cssixbasic{
S_{CS} = {1\over g_s}\,\int C^{(4)}\wedge\left({1\over 3!}F\wedge
F\wedge F\right) }
and the correction this time is parametrized as:
\eqn\wsixterm{
\Delta S_{CS} = {1\over g_s}\,\int C^{(4)}\wedge
(F\wedge W_4 + W_6) }
Here $W_4$ is given to 4-derivative order in Eq.\wfour, and $W_6$ has
been determined by explicit computation\refs\wyllard\ to be:
\eqn\wsix{
W_6 = (\tpa)^3\,{\zeta(3)\over 24 \pi^3}\, \tr (h\boldS
\wedge h\boldS \wedge h\boldS) + \ldots}
to 6-derivative order. Following the same arguments as for the 4-form
case, and restricting to the leading terms of cubic order in $F$, we
expect to find that
\eqn\sixformexp{
{1\over 3!} F\wedge F\wedge F + F\wedge W_4\Big|_{SW} + 
W_6\Big|_{SW} = {1\over 3!} \langle F\wedge F\wedge F\rangle_{*_3}}

We immediately seem to face a problem. For even integer arguments we
have the property that ${\zeta(n)\over \pi^n}$ is rational, but for
odd integer arguments there is no such property. Hence there does not
seem to be any way to obtain a number like $\zeta(3)$ by expanding
$*_3$. Fortunately, in the Seiberg-Witten limit, $W_6$ to the order
given in Eq.\wsix\ vanishes. This is because, in this limit, $h$ is
replaced by $\theta$, whose antisymmetry together with the symmetry of
$\boldS$ ensures that the trace in Eq.\wsix\ is zero. There is still
something to check, however. We have already seen that
\eqn\wfoursw{
W_4\Big|_{SW} = \half \langle F\wedge F\rangle_{*_2} - \half F\wedge F 
}
Thus Eq.\sixformexp\ implies the identity:
\eqn\fourderivid{
{1\over 3!} F\wedge F\wedge F + F\wedge\left(
\half \langle F\wedge F\rangle_{*_2} - \half F\wedge F\right)
+ W_6\Big|_{SW}
= {1\over 3!} \langle F\wedge F\wedge F\rangle_{*_3}}
which determines the Seiberg-Witten limit of $W_6$ entirely in terms
of $*_n$ products.
\eqn\wsixsw{
W_6\Big|_{SW} = {1\over 3!} \langle F\wedge F\wedge F\rangle_{*_3}
- \half F\wedge \langle F\wedge F\rangle_{*_2}
+ {1\over 3} F\wedge F\wedge F }
Since we know that the LHS vanishes to 6-derivative order, it must be
the case that the RHS is also zero to this order (in particular, the
4-derivative terms cancel out), which one can confirm by expanding
$*_3$ and $*_2$.

\subsec{8-Form Corrections, Order $F^4$}

This case is important because the explicit computation of derivative
corrections to the Chern-Simons action produces a new 8-form $W_8$,
that starts with 8 derivatives. The computed term is nonvanishing even
in the Seiberg-Witten limit. Thus we have a new numerical coefficient
and index structure to compare with the predictions of
noncommutativity. In this subsection we neglect all terms that are
higher order in $F$ compared to the leading power $F^4$.

In this case, the derivative corrections to the coupling
\eqn\cstwobasic{
{1\over g_s}\int C^{(2)}\wedge \left({1\over 4!} F\wedge F\wedge
F\wedge F\right) }
are parametrized as:
\eqn\weightterm{
\Delta S_{CS} = {1\over g_s}\,\int C^{(2)}\wedge
\left(\half F\wedge F\wedge W_4 + \half W_4\wedge W_4  + F\wedge W_6
+ W_8\right) }
Here $W_4$ and $W_6$ have already been determined, while $W_8$ has been
computed to 8-derivative order\refs\wyllard, yielding:
\eqn\weight{
W_8 = (\tpa)^4\, {\zeta(4)\over 64 \pi^4}\,\tr(
h\boldS\wedge h\boldS\wedge h\boldS\wedge h\boldS) + \ldots}
We note that $\zeta(4)={\pi^4\over 90}$ so the numerical coefficient
is indeed a rational number. Moreover, the above expression, like that 
for $W_4$, does not vanish in the Seiberg-Witten limit.

Hence repeating the arguments of the previous sections, our prediction 
is that to 4th order in $F$:
\eqn\eightformexp{
\eqalign{
{1\over 4!} F\wedge F\wedge F \wedge F 
&+ \half F\wedge F\wedge W_4\Big|_{SW} + 
F\wedge W_6\Big|_{SW} + \half W_4\wedge W_4\Big|_{SW} +
W_8\Big|_{SW}\cr
&= {1\over 4!} \langle F\wedge F\wedge F\wedge F\rangle_{*_4}\cr}}
Using Eqs.\wfoursw\ and \wsixsw\ for $W_4$ and $W_6$ in the 
Seiberg-Witten limit, we get:
\eqn\weightsw{
\eqalign{
W_8\Big|_{SW} =~& {1\over 4!}\langle F\wedge F\wedge F\wedge
F\rangle_{*_4}
- {1\over 3!} F\wedge \langle F\wedge F\wedge F\rangle_{*_3}
- {1\over 8}\langle F\wedge F\rangle_{*_2} \langle F\wedge
F\rangle_{*_2} \cr
&+ \half F\wedge F \wedge \langle F\wedge F\rangle_{*_2}
- {1\over 4} F\wedge F\wedge F\wedge F\cr }}
It is a tedious but straightforward exercise to expand the right hand
side in powers of derivatives. At the end of it, one finds that, to
8-derivative order,
\eqn\weightcheck{
W_8\Big|_{SW} = {1\over 5760}\,\tr (\theta\boldS \wedge
\theta\boldS \wedge \theta\boldS \wedge \theta\boldS) }
in perfect agreement with the Seiberg-Witten limit of Eq.\weight. Note
that this computation not only predicts the correct 8-derivative
term that appears in $W_8\Big|_{SW}$, with the correct
coefficient, but also involves a number of delicate cancellations 
between different terms on the right hand side. These cancellations
involve both 4-derivative and 8-derivative terms, and are crucial in
ensuring that the surviving 8-derivative term has precisely the index 
contractions required to match with Eq.\weight.

\subsec{4-Form Corrections, Higher Orders in $F$}

In this subsection we return to the 4-derivative, 4-form corrections
that were examined in subsection 4.1, but now we relax the requirement 
that $F$ is small. Thus we have to keep higher orders in $F$. Since the
noncommutative field strength $\hF$ is an infinite series in powers of 
$F$ and its derivatives, given by the Seiberg-Witten map, we will have 
to face this complication now. In addition, the factor $\Pf Q\over
\Pf\theta$ and the Wilson line will all make contributions. To keep
things manageable, we restrict our attention to terms involving only 4 
derivatives and work in order $F^3$.

This check is very nontrivial because, in the Chern-Simons context, it
involves for the first time all the different contributions in
Eq.\wilsoncsoper. Since two explicit $\hF$ factors are already
present, to get a third one we can expand either the Pfaffian, or the
$(1-\theta\hF)$ denominators, or the Wilson line, in each case to
first order. Also, in the second order term in $\hF$ we must insert
the Seiberg-Witten map to the lowest nontrivial order, which leads to
more $F^3$ terms.

The computation consists of adding together the following terms. For
convenience, we write out the 4-form indices $a,b,c,d$ explicitly, and
it is to be understood that they are totally antisymmetrized. The
first contribution is:
\eqn\firstterm{
\hF^2~{\rm term}:\quad
\half\langle\hF_{ab}, \hF_{cd}\rangle_{*_2} = 
\half F_{ab}F_{cd} - \theta^{ij}\langle\,
\langle A_i,\del_j F_{ab}\rangle_{*_2} -
\langle F_{ai},F_{bj}\rangle_{*_2}, F_{cd}\rangle_{*_2} }
where the Seiberg-Witten map has been inserted on the RHS. For the
rest, we get
\eqn\secondterm{
\eqalign{
&(1-\theta\hF)~{\rm denominators}:\quad
\theta^{ij}\langle\hF_{ai},\hF_{jb},\hF_{cd}\rangle_{*_3}\cr
&{\rm Pfaffian}:\quad -{1\over 4}\theta^{ij}\langle
\hF_{ji},\hF_{ab}\hF_{cd}\rangle_{*_3} \cr
&{\rm Wilson~line}:\quad
\half\theta^{ij}\del_j\langle \hA_i,\hF_{ab},\hF_{cd}\rangle_{*_3} }}
In these terms, we can replace $\hF$ by $F$ everywhere since we are
working to order $F^3$.

As a first check, it is easy to see that on replacing all $*$ products
by ordinary products, all the cubic terms add up to zero. This amounts
to the fact that there are no corrections to this Chern-Simons term that
is 0-derivative but cubic in $F$.

Now we proceed to expand the $*_2$ and $*_3$ products, keeping terms
with upto 4 derivatives. The relevant formulae are:
\eqn\relevformulae{
\eqalign{
\langle f,g \rangle_{*_2} &\sim f\,g - {1\over 24} \theta^{pr}\theta^{qs}
\del_p \del_q f\,\del_r \del_s g \cr
\langle f,g,h \rangle_{*_3} &\sim f\,g\,h - 
{1\over 24}\theta^{pr}\theta^{qs} \Big(
f\, \del_p \del_q g\,\del_r \del_s h +
g\, \del_p \del_q h\,\del_r \del_s f +
h\, \del_p \del_q f\,\del_r \del_s g \Big) }}
Using this expansion and summing up Eqs.\firstterm\ and \secondterm,
the final result for the cubic terms is then:
\eqn\csfthree{
-{1\over 12} \theta^{ij}\theta^{pr}\theta^{qs}\,
\del_p F_{ai}\,\del_q F_{bj}\,\del_r\del_s F_{cd}
+{1\over 24} \theta^{ij}\theta^{pr}\theta^{qs}\,
F_{pi}\, \del_q\del_j F_{ab}\,\del_r\del_s F_{cd}}
This is to be compared with the results of explicit computation.

From Eq.\wfour, we see that $W_4$ contains two types of $F^3$
terms. One comes from inserting the second term in Eq.\twoforms\ into
Eq.\wfour. Another arises by keeping the linear terms in Eq.\twoforms,
but noting that $h^{ij}$ in Eq.\hdef\ contains powers of $F$. In the
Seiberg-Witten limit this gives us
\eqn\hdefswlim{
\tpa h^{ij}\to \left({1\over B+F}\right)^{ij}
\sim (\theta - \theta F\theta)^{ij} +
\ldots }
The term linear in $F$ above then gives the second contribution to the
$F^3$ terms.

We therefore find that the 4-derivative contribution to $W_4$ of order
$F^3$, in the Seiberg-Witten limit, is made up of the following two
terms:
\eqn\wfourfthree{
\Big[W_4({\rm order}~ F^3)\Big]_{abcd} = {1\over 12}
\, \theta^{ij} \theta^{kl} \theta^{pq}\,\del_j\del_k F_{ab}\, 
\del_l F_{cp}\,\del_i F_{dq} 
+ {1\over 24}\theta^{ij} \theta^{kl} \theta^{pq}\,
F_{ki}\,\del_p\del_j F_{ab}\,\del_l\del_q F_{cd} }
where again we have displayed the form indices $a,b,c,d$ explicitly, and
antisymmetrization over them on the RHS is understood. Comparing
Eqs.\wfourfthree\ and \csfthree, we see after rearranging a few
indices that they agree perfectly. This once more demonstrates that the
Seiberg-Witten limit of derivative corrections in ordinary string
theory can be determined just using noncommutativity.

It should be straightforward to extend the above procedure to
4-derivative terms of order $F^4$ and higher, and compare them with
the relevant results in Ref.\refs\wyllard, though we will not do this
here.

\newsec{Conclusions}

The amazing agreement between our calculations and the boundary-state
computations performed by Wyllard\refs\wyllard\ calls for some
comment. This agreement basically stems from the fact that the
variables used in Ref.\refs\wyllard\ are, in a precise sense, the
correct ones in terms of which a comparison can be made.  Indeed, it
is a field redefinition performed in Eq.(2.15) of Ref.\refs\wyllard\
that plays the crucial role in ensuring this agreement. The motivation
for this field redefinition was to eliminate derivative corrections to
the coupling to the RR 8-form $C^{(8)}$. Because of this, the
conventional form of the Seiberg-Witten map holds for the {\it same}
choice of variables in which the results of Ref.\refs\wyllard\ are
expressed, as was seen in
Refs.\refs{\okawao,\smnvstwo,\micheliutwo}. Once this is ensured, the
variables are completely determined and there is no longer an
ambiguity of field redefinitions.

To summarize, in this paper we have demonstrated that noncommutativity
is a powerful tool in determining an infinite set of stringy $\alpha'$
corrections to the ordinary (commutative) D-brane effective action,
including couplings to closed-string backgrounds. This works basically
because the insertion of Wilson lines ensures the exact equivalence of
commutative and noncommutative actions, and because the Seiberg-Witten
limit drastically simplifies the noncommutative description while
retaining higher-derivative $\alpha'$ corrections on the commutative
side.

\medskip

\noindent{\bf Acknowledgements}

We are grateful to Sandip Trivedi for collaboration in the initial stages
of this work, and to Sandip Trivedi and  Ashoke Sen for helpful
discussions. 

\listrefs \end